\documentclass[%
 reprint,
 amsmath,amssymb,
 aps,
]{revtex4-1}

\usepackage{graphicx}
\usepackage{subfigure}
\usepackage{dcolumn}
\usepackage{bm}
\usepackage{color}

\begin{document}

\title{Statistical mechanics of dense granular fluids -\\
contacts as quasi-particles}

\author{Raphael Blumenfeld}%
 \email{rbb11@cam.ac.uk}
\homepage{http://rafi.blumenfeld.co.uk}

\affiliation{
Imperial College London, London SW7 2AZ, UK
}%
\altaffiliation{
National University of Defense Technology, Changsha 410073, Hunan, China
}%
\altaffiliation{
Cambridge University, JJ Thomson Avenue, Cambridge CB3 0HE, UK
}%


\begin{abstract}
A new first-principles statistical mechanics formulation is proposed to describe slow and dilated granular fluids, where prolonged intergranular contacts vitiate collision theory. 
The contacts, where all the important physics takes place, are regarded as quasi-particles that can appear and disappear.
A contact potential, $\nu$, is defined as a measure of the fluctuations and the mean coordination number per particle and its fluctuations are calculated as a function of it. 
This formulation extends the Edwards statistical mechanics to slow dynamic systems and converges to it when the motion stops.
The theory is applied to a model system of a simply sheared granular material in the limit of small confining stress. 
The dependence of the contact potential on the shear rate is derived, making it possible to calibrate $\nu$ experimentally and predict the coordination number distribution as a function of the shear rate. 
Setting next $\nu=0$ as the jamming point, where the critical mean coordination number is $z_c$, a finite value is found for the shear rate there, $\dot{\gamma}_c$. A simple mean field theory then yields the scaling of the mean coordination number and its fluctuations near $\dot{\gamma}_c$. Existing results in the literature appear to support the predictions. 

\end{abstract}

\maketitle


The behaviour of granular matter (GM), one of the most significant forms of matter on Earth and to human society, is still far from understood and the focus of intensive research. 
Statistical studies of static granular assemblies, dating back to the 1920s \cite{Statistics}, have been jump-started by the introduction of granular statistical mechanics (GSM) \cite{Edetal89}, which has been developed for ensembles of static systems and it is based on the entropy of structural configurations \cite{Edetal89,BlEd03,BlEd06} and of stress microstates \cite{EdBl05,Heetal07,Puetal10}. Since statistical mechanics is the most systematic method to derive equations of state and constitutive relations, extending it to dynamic systems has been a holy grail in the field. While such an extension is possible for granular gases, where collisions dominate the dynamics \cite{Go03}, attempts to formulate GSM for slowly flowing dense systems have proved difficult. In particular, attempts to model such dynamics with conventional thermal statistical mechanics have not been successful in providing general predictions. One difficulty is the inherent non-ergodicity of such systems and another is that the conventional phase space of positions and momenta is not sufficient to account for the entire entropy. The main problem is that in dense granular flows collisions are irrelevant and particles are in prolonged contacts with their neighbours. Modelling the behaviour of dense granular fluids is also useful to understanding non-brownian suspensions. 
I restrict the discussion to slow flows where an intergranular contact network can be defined and thence one can quantify the structure \cite{BlEd03,BlEd06,Bletal15} and intergranular forces \cite{Peetal06}. 
The structure and forces form the basis for statistical mechanics of static systems \cite{Edetal89, BlEd03,BlEd06,EdBl05,Heetal07,Puetal10}. Thus, not only do we need a dynamic GSM, which takes these degrees of freedom (DFs) into account but it should also converge to the, already well developed, static GSM as the dynamics slows down. The construction of such a theory is the main aim of this work. 
Since the structure and the stress state are governed by the contact network, the main idea here is to base the GSM of dynamic non-collisional systems on the contacts. Specifically, the contacts are regarded as quasi-particles that can appear and disappear during the dynamic process.

Recall that, in the static stress-structure GSM, each microstate comprises the same collection of particles and the same mean coordination number. The partition function is \cite{Bletal15}
\begin{equation}
Z_s = \int e^{-C/\tau - \chi : \mathcal{F}} \Theta d^{N_f}\vec{\phi} \ d^{N_s}\vec{\rho}
\label{ZStatic}
\end{equation}
where the vectors $\vec{\rho}$ (of length $N_s$) and $\vec{\phi}$ (of length $N_f$) consist of the structure and force DFs, respectively,
$C=\vec{\rho}\cdot A\cdot\vec{\rho}$ is a connectivity function, $A$ is a connectivity matrix,
$\tau$ is the `contacture' associated with $C$ - the analogue of the compactivity in the volume ensemble.
$\mathcal{F}$ is the force moment tensor (formed by the outer product of the intergranular forces and their position vectors, summed over particles), which couples the structure to the stress state \cite{EdBl05,Heetal07,Bletal12} and
$\chi_{ij}=1/X_{ij}$ is the inverse angoricity tensor. The function $\Theta$ includes the constraints on the systems forming the ensemble, e.g. that they are in mechanical equilibrium, generated by the same process and all have the same mean coordination number $z$. 
The weighting of the structural microstates by a connectivity function replaces the original volume function \cite{Edetal89}, which has been found to underestimate significantly the entropy \cite{Bletal15}, but this issue is tangential to the present formulation.
In general, the structural DFs should also include the parameters specifying the particle shapes. These DFs are ignored here for brevity, but could be included without loss of generality \cite{BletalOxCh15}. 

The existence of a well defined contact network establishes a quantifiable structure and its DFs \cite{BlEd03,BlEd06}, as well as enabling transmission of forces \cite{Peetal06},  which allows us to extend the functions $C$ and $\mathcal{F}$ in (\ref{ZStatic}) to slow dynamic processes. 
Additionally, $\Theta$ can be generalised to include all the same-$z$ structural and stress microstates accessible by the dynamic process. It follows that expression (\ref{ZStatic}) can describe dynamic same-$z$ structure-stress sub-ensembles. To complete the extension, we must account for the fact that the dynamics move the system between microstates with different values of ${z}$.  

The extension is best illustrated in a specific context. Consider a three-dimensional granular assembly of $N\gg 1$ convex particles of similar (but not necessarily identical) sizes, and mutual dynamic friction coefficient $\mu$. The system is a box of dimensions $W\times L\times H$, with $H$ the height between two parallel plates under a confining stress $\sigma_n$. The upper plate shears the material at a rate $L\dot{\gamma}$. 

To maintain $\sigma_n$, the system can dilate.
The internal structure and forces evolve by making and breaking contacts, whose number, ${z}N/2$, fluctuates. The dilation effects values of $z$ generically below the jamming value, ${z}_c$ and the focus here is on this regime. The value of $z_c$ depends on the dimension, the particle roughness and, to an extent, the loading history \cite{Lu16}. 
$Z_s$ can then be regarded as a sum over all the shear-induced microstates of the same total number of contacts. The general partition function of this process is constructed by summing over all the same-$z$ sub-ensembles
\begin{equation}
Z_{tot} = \sum_{k=Nz_{min}/2}^{Nz_{max}/2} Z_s e^{-\nu k}
\label{ZTotal}
\end{equation}
where $z_{min}$ ($z_{max}$) is the smallest (largest) number of contacts a particle can have. $\nu$ is a newly defined contact potential, analogous to the chemical potential in thermal statistical mechanics. 
When the motion stops and $z$ parks on a specific value, $Z_{tot}$ converges to the static one - a feature that is absent when describing the dynamics with thermal-like fluctuations and granular temperature. This statistical mechanics bridges, for the first time, between statics and dynamics in these non-ergodic systems. 

\begin{figure}
\begin{minipage}[t]{0.4\textwidth}
\includegraphics[width=0.9\textwidth]{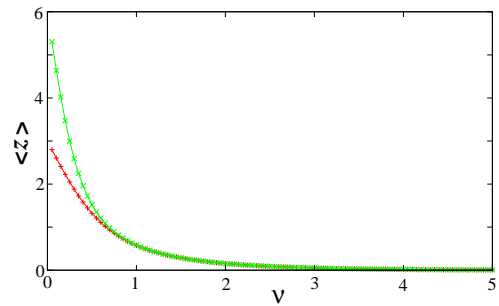}
\caption{The mean coordination number per particle as a function of the contact potential in two dimensions (red) and three (green).}
\label{fig:zav}
\end{minipage}
\end{figure}

\begin{figure}
\begin{minipage}[t]{0.4\textwidth}
\includegraphics[width=0.9\textwidth]{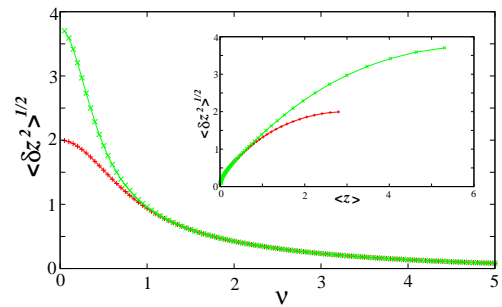}
\caption{The fluctuations of the mean coordination number per particle as a function of the contact potential in two dimensions (red) and three (green). 
Inset: $\langle \delta z^2 \rangle^{1/2}$ increases monotonically with $\langle z\rangle$ in two (red) and three (green) dimensions.}
\label{fig:dz2av}
\end{minipage}
\end{figure}

To derive explicit results, consider systems under very small confining stresses. In this limit, $\chi : \mathcal{F}\ll C/\tau$ and the latter term in $Z_s$ is negligible. 
Since only the total number of contacts is constrained, a safe assumption is that any correlations between number of contacts of different particles decays over a distance of a particle or two, and I neglect such correlations altogether.  
Additionally, it has been observed \cite{Bletal15} that, for frictional particles in this regime, $\ln{Z_s}= N\ln{Z_0}$, with $Z_0$ independent of $N$. Using this in (\ref{ZTotal}) and reordering the summation over $k$ and $N$, the sum over contacts can be decoupled from $Z_s$ and $Z_{tot}$ can be calculated  exactly
\begin{equation}
Z_{tot} = \left[ \sum_{k={z}_{min}}^{{z}_{max}} Z_0 e^{-\nu k}\right]^N = Z_s  \left[\frac{e^{-\nu {z}_{min}} - e^{-\nu({z}_{max}+1)}}{1- e^{-\nu}}\right]^N
\label{ZTotal1}
\end{equation} 
The value of ${z}_{max}$ depends on the particle size distribution, e.g., when it is very narrow, ${z}_{max}\to 6$ and $12$ in two and three dimensions, respectively \cite{Ueetal12}. It can be written as a multiple of the jamming value, $z_{max}=\kappa z_c$ ($\kappa>1$).
The value of ${z}_{min}$ depends on the dynamics: allowing or disallowing `floaters' and `rattlers' corresponds to ${z}_{min}=0$ or $2$, respectively. 
Including floaters, for illustration, we can now calculate $\langle {z} \rangle$ and $\langle \delta {z}^2 \rangle$ explicitly,
\begin{equation}
\langle {z} \rangle = \frac{1}{e^{\nu}-1} - \frac{{z}_{max}+1}{e^{({z}_{max}+1)\nu}-1}
\label{zav}
\end{equation}
\begin{equation}
\langle \delta {z}^2 \rangle = \frac{e^{\nu}}{\left(e^{\nu}-1\right)^2} - \frac{({z}_{max}+1)^2 e^{({z}_{max}+1)\nu}}{\left(e^{({z}_{max}+1)\nu}-1\right)^2}
\label{dz2av}
\end{equation}
These are plotted in figures \ref{fig:zav} and \ref{fig:dz2av}. 
Relations (\ref{zav}) and (\ref{dz2av}) provide a way to estimate the contact potential from the steady-state time series of ${z}$.

Increasing the shear rate should increase particle velocities and therefore the rates of contact creation and annihilation. One then expects $\partial\nu/\partial\dot{\gamma}>0$. 
To test this and derive experimental predictions, the next aim is to relate $\nu$, $\langle {z} \rangle$, $\langle \delta {z}^2 \rangle$ and $\dot{\gamma}$.

At steady state, energy is dissipated through particles sliding over one another against friction, $\dot{E}_{fric}$, and contact making, $\dot{E}_{clap}$.  
A contact is made when two particles collide without being able to recoil. Such a `clap' converts kinetic energy to density waves in the particles and in the surrounding fluid. 
The dissipation is balanced by the energy pumped into the system by the shear.
\begin{figure}
\begin{minipage}[t]{0.4\textwidth}
\includegraphics[width=0.9\textwidth]{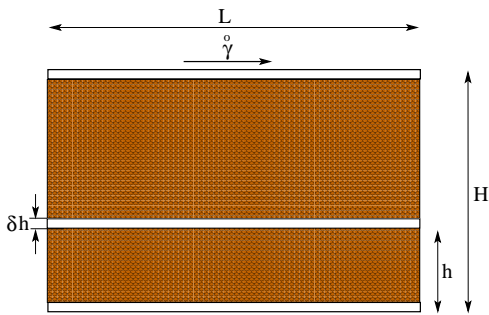}
\caption{A sketch of the sheared system.}
\label{fig:System}
\end{minipage}
\end{figure} 

Consider an imaginary sheet of thickness $\delta h$ ($\leq$ particle size), parallel to the shear plates, $h$ away from the bottom plate (figure \ref{fig:System}). The sheet contains, on average, $nWL$ particles, each having at most one contact within the sheet. The mean number of contacts per sheet is then 
$nWL/\langle z\rangle$, of which only a fraction $\Delta$ are newly made within a time interval $\delta t$. 
The steady-state contact creation, $\dot{z}_+$, and annihilation, $\dot{z}_-$, rates are $\dot{z}_+ = \dot{z}_- \approx \langle \delta z^2 \rangle^{1/2}$ and, per contact, 
$\Delta = \dot{z}_+/ \langle z \rangle \approx  \langle \delta z^2 \rangle^{1/2}/\langle z \rangle$. 
The average normal force transmitted through the sheet per contact is $\sigma_n WL /(nWL/\langle z\rangle)$. 
Only a fraction $p$ (of order $3\%$ \cite{Zh03}) of all the $(1-\Delta)nWL/\langle z\rangle$ contacts, existing prior to this $\delta t$, slide against friction \cite{SlidCont}. Thus, the energy dissipated by friction within the sheet is $\delta\dot{E}_{fric} = p\mu \sigma_n W L(1-\Delta) v(h) \delta h$. Assuming no shear banding, we can approximate the velocity profile by $v(h)=L\dot{\gamma}h/H$ and integrate over $h$ to find the total rate of energy dissipated by this mechanism
\begin{equation}
\dot{E}_{fric} = \frac{\mu p\sigma_n V}{2}\left( 1 - \Delta \right)L\dot{\gamma}
\label{EFriction}
\end{equation}
An exponential form of $v(h)$ would yield a similar result. 

The energy lost in contact making is primarily kinetic. 
In the sheet's moving frame, the particles' relative velocities are of order $\langle \delta v^2\rangle^{1/2}$, leading to a clapping energy loss of roughly $m\epsilon \langle \delta v^2\rangle$ per clap, where $m$ and $\epsilon$ are, respectively, a typical particle mass and the restitution coefficient. The number of newly made contacts is the number of particles in the sheet, $nWL\delta h$, times the probability that two neighbours are in contact, $\langle z\rangle/z_{max}$, times the probability that the contact has been made in $\delta t$, $\Delta$.
Assuming $\langle \delta v^2\rangle\sim \left(L\dot{\gamma}\right)^2 h/H$, with a proportionality constant of order 1 \cite{Meetal14} and integrating over $h$, yields the clapping energy dissipation rate,
\begin{equation}
\dot{E}_{clap} = \frac{\rho \epsilon V\Delta}{2z_{max}} L^2\dot{\gamma}^2
\label{EClap}
\end{equation}
with $\rho=mn$ the mass density. Dependence of $n$ on $h$, due to dilation, would affect this result negligibly. Similarly, a different dependence of $\langle \delta v^2\rangle$ on $h$ would change this result by a factor of order 1, at most.

\begin{figure}
\begin{minipage}[t]{0.4\textwidth}
\includegraphics[width=0.9\textwidth]{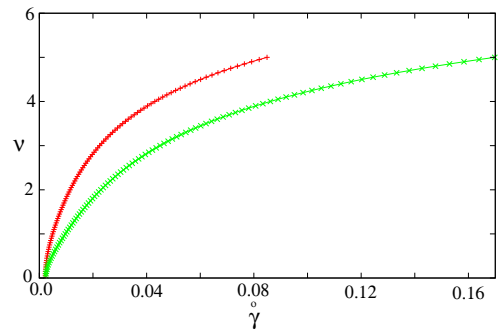}
\caption{The contact potential as a function of the shear rate in two (red) and three (green) dimensions.}
\label{fig:NuGammadot}
\end{minipage}
\end{figure}

\begin{figure}
\begin{minipage}[t]{0.4\textwidth}
\includegraphics[width=0.9\textwidth]{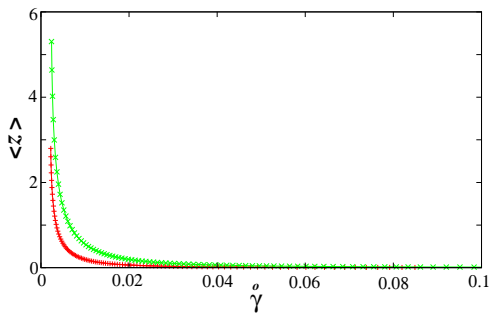}
\caption{The mean coordination number per particle as a function of the the shear rate in two (red) and three (green) dimensions.}
\label{fig:ZvsGammadot}
\end{minipage}
\end{figure}

\begin{figure}
\begin{minipage}[t]{0.4\textwidth}
\includegraphics[width=0.9\textwidth]{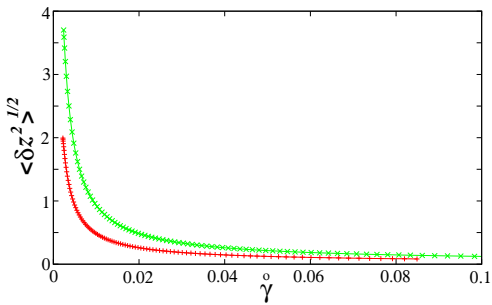}
\caption{The fluctuations of the mean coordination number per particle, $\langle \delta z^2 \rangle^{1/2}$, as a function of the shear rate in two (red) and three (green) dimensions.}
\label{fig:dZvsGammadot}
\end{minipage}
\end{figure}

The power pumped into the system is the product of the top plate velocity, $L\dot{\gamma}$, and the friction force on it, $\mu_H\sigma_n WL$, where $\mu_H$ is the ratio of the shear to normal stresses at $H$. Equating it to the total dissipation rate, (\ref{EFriction}) and (\ref{EClap}), and rearranging, we obtain,
\begin{equation}
\dot{\gamma} = \frac{2 \sigma_n z_{max}}{\rho \epsilon L}\frac{\mu_H - \mu p\left(1 - \Delta \right)}{\Delta\langle z\rangle}
\label{EBalance}
\end{equation}
This expression gives a novel prediction on the dependence of features of the contact number distribution on the shear rate. Since $\langle z\rangle$ and $\Delta$ are known functions of $\nu$ from (\ref{zav}) and (\ref{dz2av}), then this expression provides another way to estimate the contact potential and calibrate it against the shear rate. 
Figures \ref{fig:NuGammadot} - \ref{fig:dZvsGammadot} show these relations for three-dimensional granular systems, sand particles density realistic values for $\mu_H$ \cite{Zh03,GuMo04,RaRo03}. Included in these figures are plots for two-dimensional systems with the same parameters.
As expected, $\nu$ increases with $\dot{\gamma}$ and both $\langle z\rangle$ and $\langle \delta z^2 \rangle$ decrease with it.

Existing observations seem to validate these results. A monotonic drop in $\langle z\rangle$, and consequently in the packing fraction, as $\dot{\gamma}$ increases under constant confining stress, has been observed widely, e.g. in \cite{Zh03,Cretal05,Boetal11,WyCa14}. While the variation of $\langle \delta z^2 \rangle^{1/2}$ with $\dot{\gamma}$ during simple shear is less documented, the increase of $\langle \delta z^2 \rangle^{1/2}$ with $\langle z\rangle$, predicted here (inset in figure \ref{fig:dz2av}), has been observed experimentally in random packs of spheres \cite{Od77}. 


It is tempting to identify $\nu=0$ as the jamming point by setting $\kappa$ so that $\langle {z}(\nu=0) \rangle = {z}_c$. 
The value of $\kappa$ depends on the system and on whether floaters and rattlers are included or not. For the above case, $\kappa=2$. 

Interestingly, the shear rate at the jamming point is finite, 
$\dot{\gamma}_0=\frac{4\sigma_n}{\rho\epsilon L}\left[\sqrt{\frac{3}{z_c(z_c+1)}}\left(\mu_H - p\mu\right) + p\mu\right]$, and to linear order in $\nu$
\begin{eqnarray}
\delta\dot{\gamma} & = & \dot{\gamma}-\dot{\gamma}_0 = \frac{4 \sigma_n p\mu \left(z_c+1\right)}{3\rho \epsilon L}\nu \label{JamPoint0}\\
z_c - \langle {z} \rangle & = & \frac{z_c(z_c+1)}{3}\nu = \frac{\rho\epsilon L}{4\sigma_n p\mu} \frac{z_c(z_c+1)}{z_c+4}\delta\dot{\gamma}
\label{JamPoint1}
\end{eqnarray}
Again, the dependence of the density, $\rho$, on $\langle {z} \rangle$ in this expression, due to dilation, is negligible.
This behaviour is supported qualitatively by simulations \cite{Peetal06,Cretal05,LuPri}.


{\it Conclusion.} -- A first-principles statistical mechanical theory has been formulated for slow dense granular fluids. It is based on treating the contacts as quasi-particles that can appear and disappear as the material evolves. The contact number fluctuations are characterised by a contact potential, $\nu$, analogous to the chemical potential in thermal statistical mechanics. In relatively uniform systems, e.g. in the absence of shear banding, regions with high mean coordination number dilate and `shed' contacts, while low contact regions densify to gain some - the rates of which are controlled by $\nu$. 

Applying the formalism to simple shear under constant confining pressure, explicit relations have been derived for the dependence of the expected mean coordination number, $\langle{z}\rangle$, and its fluctuations, $\langle\delta^2{z}\rangle$, on $\nu$. 
The contact potential has been then related explicitly to the shear rate at steady state, using balance between the shearing energy and the energy dissipated by particle friction and clapping. This enabled relating $\langle{z}\rangle$ and $\langle\delta^2{z}\rangle$ to the shear rate, making it possible to calibrate the contact potential in shear experiments. The cited literature supports well relation $\langle{z}\left(\dot{\gamma}\right)\rangle$ and, tentatively, the dependence of ratio 
$\langle\delta^2{z}\rangle^{1/2}/\langle{z}\rangle$ on the strain rate, but no systematic results in this regime exist for $\langle\delta^2{z}\left(\dot{\gamma}\right)\rangle$. 
Note that since $p\ll 1$ \cite{Peetal06,Zh03,Cretal05} then (\ref{EBalance}) yields $\langle\delta^2{z}\rangle^{1/2}\sim\dot{\gamma}^{-1} + O(p)$, which could be tested experimentally.
It would be useful to test these results further by simple shear experiments in the steady state regime, e.g., using a continuous parallel plate shear apparatus \cite{ShearStadium}. 
We are looking forward to experiments testing the concept of the contact potential, which could also be done in non-Brownian dense suspensions \cite{WyCa14}.

It would also be useful to test the assumptions underlying the clapping energy dissipation, e.g. by acoustic emission experiments \cite{AE}, and whether $E_{clap}\sim\dot{\gamma}^2$. This relation has been examined numerically in \cite{He13}, who find that the total dissipation appears to be proportional to $\dot{\gamma}^2$. However, it is likely that the effect of sliding friction in those simulations is too small due to the very low fraction of mobilised sliding contacts.

Conjecturing that $\nu=0$ should correspond to the jamming point enables a derivation of the finite value of the shear rate, at which jamming occurs, as well as the
scaling of $z_c-\langle {z} \rangle$ with shear rate near jamming in the mean-field approximation, which this analysis is based on. The analysis can be readily extended to the dependence of any moment of the distribution of $z$ on the shear rate. This author plans to apply the formalism to other dense granular flows to test the generality of this approach.

\begin{acknowledgments}
Thanks go to D. Frenkel and S. Amitai for comments and to S. Luding for discussions. 

\end{acknowledgments}

\nocite{*}


\end{document}